\documentclass{article}
\usepackage{ijcai25}

\usepackage{fontawesome5}
\usepackage{enumitem}
\usepackage{times}
\usepackage{soul}
\usepackage{url}
\usepackage[hidelinks]{hyperref}
\usepackage[utf8]{inputenc}
\usepackage[small]{caption}
\usepackage{graphicx}
\usepackage{amsmath}
\usepackage{natbib}
\usepackage{amsthm}
\usepackage{booktabs}
\usepackage{algorithm}
\usepackage{algorithmic}
\usepackage[switch]{lineno}
\usepackage{txfonts}
\usepackage[table,xcdraw]{xcolor} 
\usepackage{amsmath} 
\usepackage{amsfonts} 
\usepackage{amssymb} 
\usepackage{geometry} 
\usepackage{tikz}

\title{Inhibiting Alzheimer’s Disease by Targeting Aggregation of $\beta$-Amyloid}

\author{
Ananya Joshi$^1$
\and
George Khoury$^1$\and
Christodoulas Floudas$^1$
\affiliations
$^1$Princeton University\\
}

\begin{document}

\maketitle

\begin{abstract}
Alzheimer's disease is characterized by dangerous amyloid plaques formed by deposits of the protein $\beta$-Amyloid aggregates in the brain. The specific amino acid sequence that is responsible for the aggregates of $\beta$-Amyloid is lys-leu-val-phe-phe (KLVFF). KLVFF aggregation inhibitors, which we design in this paper, prevent KLVFF from binding with itself to form oligomers or fibrils (and eventually plaques) that cause neuronal death. Our binder-blocker peptides are designed such that, on one side, they bind strongly to KLVFF, and on the other side, they disrupt critical interactions, thus preventing aggregation. Our methods use optimization techniques and molecular simulations and identify 10 candidate sequences for trial of the 3.2 million possible sequences. This approach for inhibitor identification can be generalized to other diseases characterized by protein aggregation, such as Parkinson's, Huntington's, and prion diseases.

\end{abstract}

\maketitle
\section{Introduction}

\textit{This manuscript was originally written in November 2014 and is being uploaded for archival purposes. Please note that this version reflects the state of the field as of November 2014, and it has not been updated to incorporate newer results or methods published since then.}

Dementia is a neurodegenerative disease that currently affects more than 44 million people globally, a number that is projected to triple by 2050 \citep{1}. The most common occurrence of dementia is Alzheimer’s Disease (60-80\% of cases). Alzheimer’s Disease (AD) is a slowly progressive disease, which suggests, in many cases, that the current treatments come too late and are ineffective \citep{2}. 

Protein aggregation plays a central role in the progress of AD \citep{3}. One key pathogen of AD is a buildup, or aggregation, of $\beta$-Amyloid plaques in the brain. During the aggregation process, $\beta$-Amyloid bind with with other $\beta$-Amyloid. As aggregation progresses from dimers, to soluble oligomers that bind to nerver cell recepters and can lead to neuronal death, or plaques. 

Targeting the $\ beta$-amyloid aggregation process at its initiation is more promising than targeting the aggregates, as there is no definitive kinetic pathway for AD aggregation \citep{1}. The specific self-aggregating sequence of amino acid residues responsible for the joining, or aggregation process, is lys-leu-val-phe-phe (KLVFF) \citep{4}. Preventing these sequences from aggregating is promising in preventing amyloid pathogenesis and fibrillogenesis \citep{2}, supporting that they could be useful at different stages of the disease. Additionally: 

\begin{enumerate}
    \item Peptide inhibitors can be readily modified for chemical stability \citep{4}.
    \item Peptide-based drugs generally have higher specificity and fewer side effects \citep{5} 
    \item Peptide-based drugs are small enough to transport easily through the human body. 
\end{enumerate}

These properties give peptide inhibitors an advantage over other compound or element-based inhibitors. Still, designing an appropriate peptide inhibitor must satisfy several constraints: 

\begin{enumerate}
    \item High fold specificity, or the ability of the peptide to fold into a given structure 
    \item Favorable binding energy, which shows how well the peptide binds to KLVFF 
    \item Satisfaction of natural constraints- interacting beneficially with the human body
\end{enumerate}

With 20 natural amino acids, a 5-residue peptide sequence (e.g., KLVFF) can be formed in 3.2 million (20 $\hat{}$ 5) ways. This research identifies candidate amino acid residue sequences that meet literature-based constraints to provide optimal candidates for creating peptide inhibitors with stable interactions at the molecular level. In the past, the discovery of these inhibitors was described as “often fortuitous rather than rational”\citep{4}. However, in this research, inhibitory peptides are identified systematically through computational protein design, which involves combinatorial optimization, molecular modeling, and virtual screening. 

Specifically, combinatorial optimization methods are used to obtain optimal solutions in the face of an excessive number of possible solutions and strict constraints, like finding optimal inhibitors of  $\beta$-Amyloid from 3.2 million possibilities \citep{6}. The objective of the optimization here is to find inhibitory peptide sequences that maximize fold specificity, minimize binding energy and fit natural biological constraints. To do so, the candidate solutions from the optimization stage are subjected to all-atom molecular dynamics simulations to find a predictive estimate of the candidate inhibitor’s binding affinity to inhibit the key interactions involved in AD amyloidogenesis. 

This procedure can be used for other neurodegenerative and aggregation based diseases, like Type II Diabetes and prion diseases, with flexible/weak NMR templates that could not be previously evaluated.

\subsection{Developing a Model}: A structural trajectory model for a 40-Residue $\beta$-Amyloid Fibril with Two-Fold Symmetry, Positive Stagger (PDB 2LMN where 2LMN is the Protein Data Bank code of this structure), consists of 10 NMR-derived states that show how  $\beta$-Amyloid  proteins interact with other $\beta$-Amyloid proteins flexibly \citep{8}. This base model takes into account key secondary and tertiary structures. As in AD progression, the KLVFF sequence in this model self-aggregates by finding other KLVFF sequences in different $\beta$-Amyloid proteins to form a plaque. An inhibitory sequence that binds to KLVFF on one end and disrupts binding on the other end would prevent $\beta$-Amyloid proteins from self-aggregating.

\begin{figure}[h]
\caption{Edited model for Chain C with Chain C highlighted in red (visualization using PyMol).}
\centering
\label{fig1}
\includegraphics[width=0.5\textwidth]{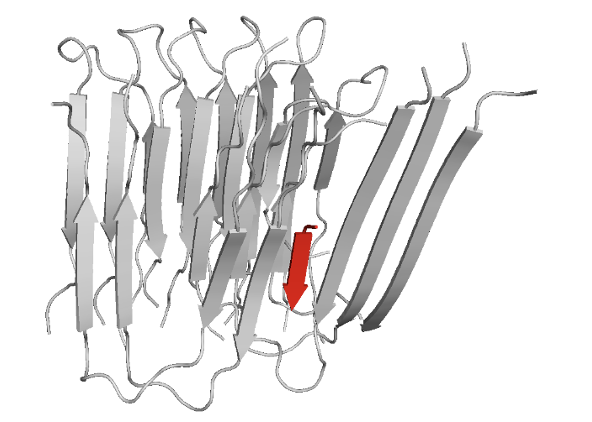}
\end{figure}

In each of the 10 NMR derived states, there are 12 different chains labeled A through L. One chain is equivalent to one $\beta$-Amyloid  protein. For every NMR-derived state, the KLVFF sequence is isolated in chain C, allowing new sequences to be based on the KLVFF backbone. The isolated sequence was blocked with neutral compounds (ACE and NHE) to prevent spurious end-effects. As seen in Figure \ref{fig1}, the model contains the isolated KLVFF sequence in Chain C, while all the other chains remain intact. 

This modified model shows that the KLVFF sequence from chain C aggregates with the KLVFF occurrences in the adjacent chains (B and D). 

\subsection{Optimization Overview}: Figure 2 shows how the top 20 sequences were determined from the 3.2 million sequences using the integrated computational framework developed in this study using Python.

\begin{figure}[h]
\caption{Python integration framework for modeling and simulation.}
\centering
\label{fig2}
\includegraphics[width=0.5\textwidth]{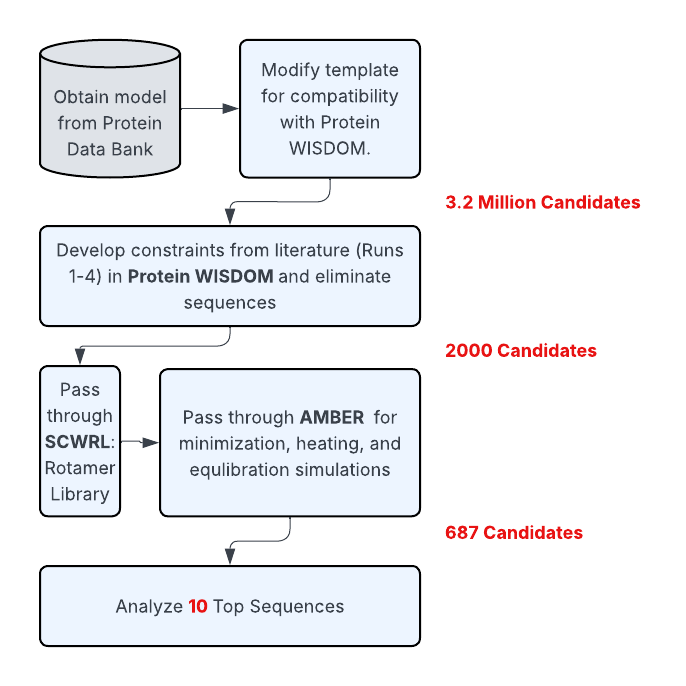}
\end{figure}

Two thousand candidate sequences remained after using mutational and biological constraints in Protein WISDOM \citep{9}. Protein WISDOM uses a combinatorial global optimization method, the branch and bound technique, to solve the optimization problem defined in Equation Set 1.  

\begin{align*}
\min_{y^j_i, y^l_k} & \sum_{i=1}^n \sum_{j=1}^{m_i} \sum_{k=i+1}^n \sum_{l=1}^{m_k} \sum_{d:disbin(x_i,x_kd)}^{b_m} E^{jl}_{ik}(x_i,x_k)b_{ikd}w^{jl}_{ik} \\
\text{subject to} \\
& \sum_{j=1}^{m_i} y^j_i = 1 \quad \forall i \\
& \sum_{j=1}^{m_i} w^{jl}_{ik} = y^l_k \quad \forall i, k > i, l \\
& \sum_{l=1}^{m_k} w^{jl}_{ik} = y^j_i \quad \forall i, k > i, j \\
& \sum_{l=1}^{m_k} w^{jl}_{ik} = y^j_i \quad \forall i, k > i, j \\
& \sum_{d:disbin(x_i,x_kd)}^{b_m} b_{ikd} = 1 \quad \forall i, k > i \\
& y^j_i, y^l_k, w^{jl}_{ik}, b_{ikd} \in \{0,1\} \quad \forall i, j, k > i, l, d
\end{align*}
\centerline{Equation set 1}

The first line of Equation Set 1 defines the objective function in terms of potential energy, and the rest of the equation provides structural constraints in terms of integer variables. The final design used the high Resolution Centroid-Centroid 8 Bin Potential for the Distance-Dependent Force Field. It sums potential energies based on distances between the centroids of each amino acid \citep{9} used in the objective function, making it a mixed integer programming problem.

Selected sequences are then run through a fold specificity stage and a binding affinity stage. Fold specificity is a measure of how well the designed sequence adopts the template fold. Equation set 2 defines fold specificity calculated from Boltzmann distributions related to conformers experienced.

\begin{equation*}
f_{spec} = \frac{\sum_{i \in \text{novel}} e^{-\beta E_i}}{\sum_{i \in \text{native}} e^{-\beta E_i}}
\end{equation*}
\centerline{Equation set 2}

Approximate binding affinity, a measure of how well the designed sequence binds to a given protein target, is shown in equation set 3. 

\begin{equation*}
K^* = \frac{q_{PL}}{q_P q_L}
\end{equation*}

\begin{align*}
q_{PL} = \sum_{b \in B} e^{-\frac{E_b}{RT}}, q_P = \sum_{f \in F} e^{-\frac{E_f}{RT}}, q_{PL} = \sum_{l \in L} e^{-\frac{E_l}{RT}}
\end{align*}
\centerline{Equation set 3}

Equation sets listed are from the Smadbeck et al. paper \citep{9}. These steps resulted in the two thousand output sequences 

Then, four sets of unique constraints were inputted and run through the Protein WISDOM web server. Each run, detailed by the rules in Table 1, varied in mutation sets and biological constraints. These were based on the properties of KLVFF that were consistent across multiple papers and previous sequences that were compatible in human bodies \citep{3, 4, 10, 15, 16, 17, 18, 19, 20, 21, 22, 23}.

\begin{table*}[htbp]
\centering
\label{tab:peptide_design_rules}

\begin{tabular}{p{0.23\linewidth}|p{0.23\linewidth}|p{0.23\linewidth}|p{0.23\linewidth}}
\multicolumn{4}{p{\linewidth}}{\textbf{Inputs: } {Modified 2LMN structure from Base Model, Mutation Sets, Biological Constraints}} \\
\hline
\textbf{Rule 1:} & \textbf{Rule 2:} & \textbf{Rule 3:} & \textbf{Rule 4:} \\
\hline
\begin{itemize}
    \item Charged first residue at neutral pH positive / basic
    \item First position only K or R (not H)
    \item Other residues constrained by SASA
\end{itemize}
&
\begin{itemize}
    \item Free first residue
    \item Overall +1 charge
\end{itemize}
&
\begin{itemize}
    \item Free first residue
    \item Overall neutral charge
\end{itemize}
&
\begin{itemize}
    \item Charged first residue at neutral pH negative / acidic
    \item First position only D or E (not C)
    \item Other residues constrained by SASA
\end{itemize} \\
\multicolumn{4}{c}{
\vspace{0.1cm} 
\begin{tikzpicture}[>=latex, shorten >=1pt, shorten <=1pt]
\draw[->,line width=1.5pt] (0.1,0.2) -- (0.1,-0.2); 
\draw[->,line width=1.5pt] (3.1,0.2) -- (3.1,-0.2); 
\draw[->,line width=1.5pt] (6.1,0.2) -- (6.1,-0.2); 
\draw[->,line width=1.5pt,teal] (9.1,0.2) -- (9.1,-0.2); 
\end{tikzpicture}
\vspace{0.1cm} 
}\\
\hline
\multicolumn{3}{p{0.69\linewidth}|}{\centering {Blocked at the N-terminus}} & {Blocked at the C-terminus with an Acetyl to reduce the number of hydrogen bonds with water.} \\
\hline
\end{tabular}
\caption{The four different runs with constraints detailed based on: \citep{3, 4, 10, 15, 16, 17, 18, 19, 20, 21, 22, 23}.}
\end{table*}

Each of the 4 runs produced 500 sequences that met these constraints. Of the resulting unique sequences, sequences that had poor potential energies or were known to self-aggregate were eliminated. Additionally, patterns observed in sequences that previously failed from prior literature were used to narrow down from the 2000 sequences to 687 sequences.

\footnote{It should be noted that the isolated KLVFF sequence (from Chain C), which is being modified to create peptides, is a ligand. The receptor refers to all parts of the trajectory except the ligand. The complex is the ligand plus the receptor.}

\begin{figure}[h]
\caption{PyMol visualiztion using a sample of random sequences overlayed so that the side chains a revisible (left), but all of the sequences have the same backbone (right) }
\centering
\label{fig3}
\includegraphics[width=0.5\textwidth]{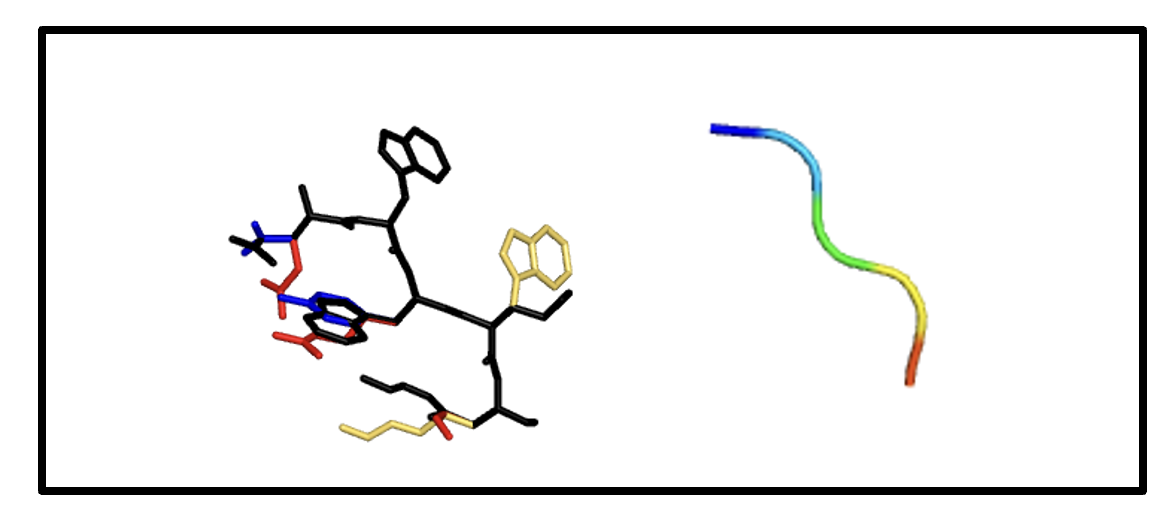}
\end{figure}

Then, initial models of each modified peptide chain were created through using the SCWRL rotamer library \citep{11}, assuming a common initial backbone, as seen in Figure \ref{fig3} \footnote{This assumption can be overcome during the simulations if necessary.}. 

The next step in finding candidate inhibitors is to run molecular dynamics simulations on these top 687 sequences. Molecular Dynamics (MD) is a simulation technique that determines the trajectories and interactions of atoms by solving for Newton's motion equations after atoms are allowed to interact at each instant in time, as defined by force fields \citep{12}. This method uses the AMBER (Assisted Model Building with Energy Refinement) software for molecular dynamic simulations and was used to obtain estimates of the binding energy of these 687 sequences \citep{13}. 

Each ligand was then determined, run through simulation, and evaluated in each complex. Because each ligand changes, new protein files for each modified C chain were created. Then, initial topology files were generated by using the force field 03 model (ff03), which dampens charges on the protein so that the charges are realistic \textit{in situ} rather than derived from the gas phase \citep{14}. This was followed by Molecular Dynamics simulations, which utilize optimization to perform minimization, heating, and equilibration of the protein. Finally, the binding energies for these remaining sequences were ranked and evaluated.

\section{Results and Discussion}

\begin{figure}[h]
\caption{Binding energy and $\delta$ Binding energy for all 687 sequences.}
\centering
\label{fig4}
\includegraphics[width=0.5\textwidth]{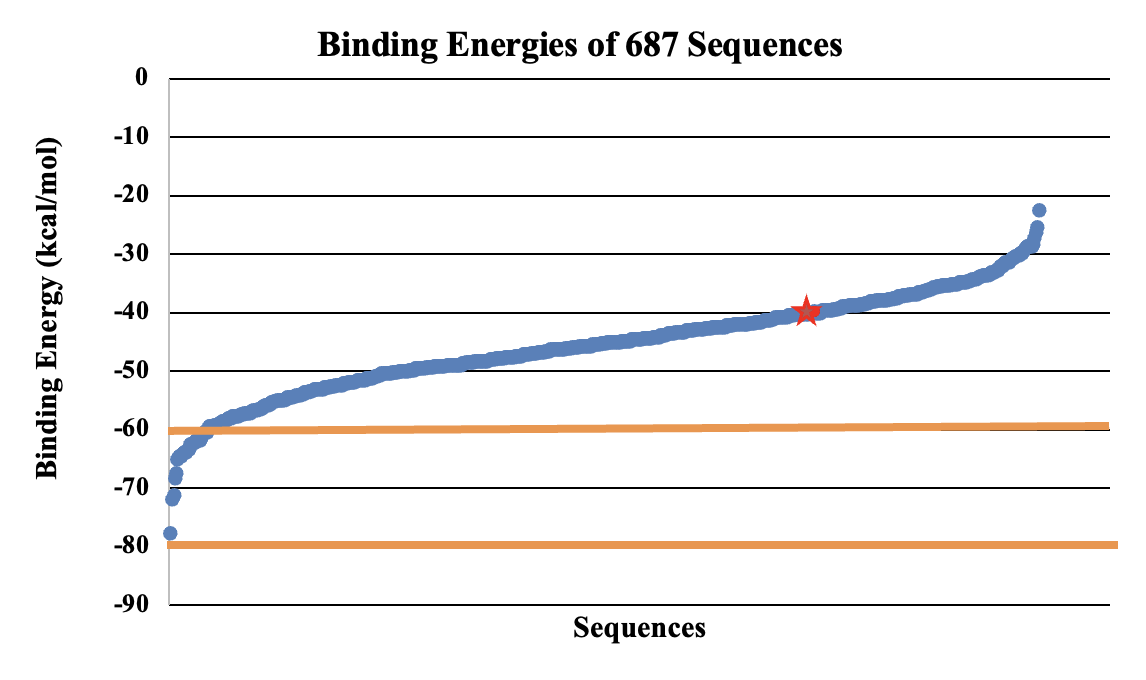}
\end{figure}

Figure \ref{fig4} presents the distribution of binding energy for all the 687 sequences, with the red mark representing the binding energy of the KLVFF sequence itself. 

\begin{figure}[h]
\caption{Bar Graph comparing Sequences (ranked by Binding Energy in kcal/mol) to native (binding energy of KLVFF to itself) with error bars}
\centering
\label{fig5}
\includegraphics[width=0.5\textwidth]{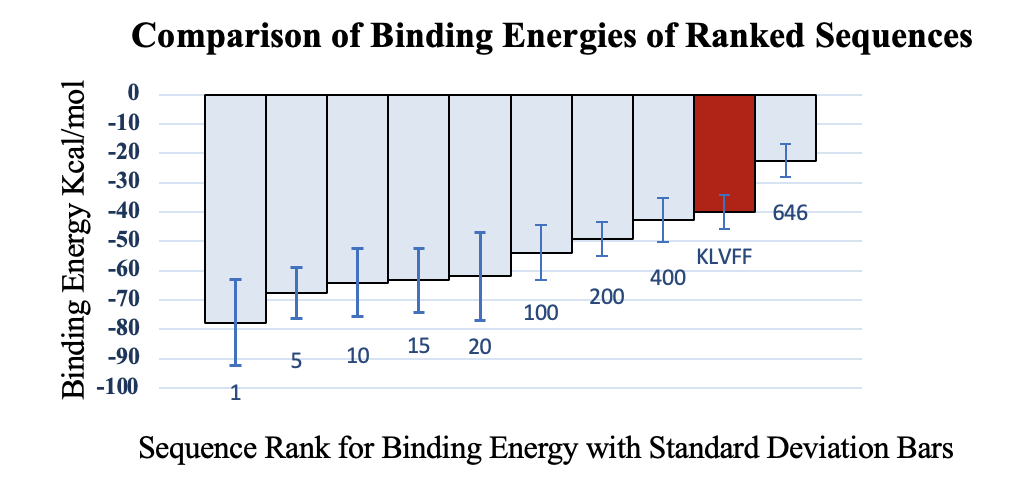}
\end{figure}

Out of the 687 sequences that were run through AMBER simulation, only a handful had very low Binding Energies. As more experimental research is pursued, it is important to consider all sequences with binding energies statistically significantly better than KLVFF as open options (see Figure \ref{fig5}). With current literature, sequences with energies between the -60 and -80 range kcal/mol are ideal and would be the most potent. 

Top ranked sequences included frequent occurrences of Tryptophans (W) in the results. Tryptophan has very favorable binding energy due to its ability to form large non-polar interactions, but it usually has poor fold specificity. This means that it may not be viable in the human body due to being too ``sticky" \citep{24}. Thus, sequences with more than three W's in a row were eliminated. 

Ultimately, 10 sequences were identified and compared with the native binding energy of KLVFF, which was -40 kcal/mol. The $\delta$ binding energy is simply the binding energy of the sequence – binding of the native (KLVFF).

Figure \ref{fig5} presents these top ten sequences with their binding energies. The figure also shows the error bars related to each sequence. In this image, it can be seen sequence with the best binding free energy is the most promising sequence because of the high binding strength relative to KLVFF.

\section{Conclusion}

This research creates a procedure and an integrated framework to screen candidate inhibitors designed through an advanced combinatorial optimization engine for flexible templates, and identified promising candidate sequences for peptide inhibitors of $\beta$-Amyloid.

First, this work established a novel procedure and integrated framework for a flexible template which can be applied in finding inhibitor sequences for other neurodegenerative diseases and in binding applications in bio-material engineering.

Second, 10 optimal candidates of peptide inhibitors for Alzheimer's were identified. These sequences are candidates for further experiments.

This procedure is consistent with all the metrics initially established: fold specificity, binding energy, and possibility of positive human interaction. Some sequences identified systematically demonstrated previously established desirable physical properties, such as the presence of alternating polar and non-polar amino acids, and the other sequences have not yet been tested. In fact, in the Kokkoni et al. paper, there was one sequence that matched the sequences generated within this study \citep{10}, which shows that this procedure independently found a candidate sequence that had previously been successful in inhibiting amyloid formation.   

Experimental considerations and possible improvement strategies include investigating bioavailability, systematically filtering sequences that self-aggregate, studying peptide access to the desired site of action,  the D chirality of these sequences to increase half-life and metabolic stability, and the N-methylation in the 5th position \citep{10, 16}.

\bibliographystyle{named}
\bibliography{ijcai25}

\end{document}